\documentclass[pre,twocolumn,aps,showpacs,superscriptaddress]{revtex4}

\usepackage{color,psfig,graphics,psfrag,amsmath,amssymb,amsfonts,rotating}

\begin{document}

\title{Localization threshold of Instantaneous Normal Modes from
level-spacing statistics}


\author{Stefano Ciliberti}

\affiliation{Departamento de F\'{\i}sica Te\'orica I, Universidad
Complutense de Madrid, Madrid 28040, Spain}

\author{Tom\'as S.~Grigera}

\affiliation{Instituto de Investigaciones Fisicoqu\'\i{}micas
Te\'oricas y Aplicadas (INIFTA, CONICET--UNLP), c.c. 16, suc. 4, 1900
La Plata, Argentina.}


\begin{abstract}
We study the statistics of level-spacing of Instantaneous Normal Modes
in a supercooled liquid. A detailed analysis allows to determine the
mobility edge separating extended and localized modes in the negative
tail of the density of states. We find that at temperature below the
mode coupling temperature only a very small fraction of negative
eigenmodes are localized.
\end{abstract}


\maketitle

\section{Introduction}

The Instantaneous Normal Modes (INM) of a liquid are the eigenvectors
of the Hessian (second derivative) matrix of the potential energy,
evaluated at an instantaneous configuration. The interest in the
equilibrium-average properties of the INM originates in the proposal
\cite{keyes} to use them to study liquid dynamical properties,
especially diffusion, which is considered to be linked to unstable
modes (a subset of the modes with negative eigenvalue) \cite{BeLa,
ScioTa97, donati00, lanave01, deo, lanave02}. They have been naturally
applied to address the problem of the glass transition: the glass
phase is viewed as that where free ({\sl i.e.\/} non-activated)
diffusion is absent, and the disappearance of diffusion should be
linked to that of the unstable modes \cite{BeLa}. The identification
of these unstable modes presents some problems
\cite{critique,response}, and the localization properties of the INM
are of interest. This is the problem we address here.

Localization is an interesting and difficult problem in its own
right. Given an $N\times N$ random matrix defined by the probability
distribution of its elements in some (typically site) base, the
problem is to determine whether an eigenmode projects to an extensive
number of base vectors (extended state) or not (localized state) in
the large $N$ limit. Only in a few cases there is a theoretical
solution for this problem~\cite{Lee, anderson, bouchaud}. From the
point of view of random matrix theory, the INM are the eigenvectors of
an Euclidean random matrix (ERM)~\cite{mepaze}, and the problem of
localization in ERMs has been recently addressed analytically in the
dilute limit~\cite{noi}. Clearly, the problem is also hard from the
numerical point of view, involving a question about the thermodynamic
limit. Quantities such as the participation ratio~\cite{bede}, which
can distinguish between localized and extended states but require
knowledge of the eigenvectors, are problematic numerically because
computation of eigenvectors is very expensive for large systems.

Here we explore an approach~\cite{carpena} relying only on
eigenvalues, based on the fact that the statistics of level spacing is
strongly correlated with the nature of the eigenmodes (see
sec.~II). We apply it for the first time to a model glass-forming
liquid at a temperature below the mode coupling temperature $T_c$
\cite{goetze}. Our work can be regarded as an extension of the results
of ref.~\onlinecite{deo}, as far as we perform a detailed analysis of
the level spacing. Our emphasis is on the exploring the usefulness of
level-spacing statistics as a means to obtain a localization threshold
in off-lattice systems and the possible limitations of this technique.


\section{Theoretical background}

The spectral function of the eigenvalues of a random matrix is
$S(\lambda) = \sum_i \delta(\lambda-\lambda_i)/N$. Its (disorder-)
average is the density of states $g(\lambda)$ (DOS). The cumulative
spectral function for this particular realization of the disorder,
\begin{equation}
  \eta(\lambda)
  =
  \int_{-\infty}^{\lambda} d\lambda' \, S(\lambda')
  =
  \frac 1N \sum_{i=1}^N \theta(\lambda-\lambda_i) ,
  \label{step}
\end{equation}
can be decomposed into a smooth part plus a fluctuating term
$\eta_\mathrm{fluct}(\lambda)$, whose average is zero. The smooth part
is then
\begin{equation}
  \zeta(\lambda)\equiv\langle \eta(\lambda)\rangle 
  = 
  \int_{-\infty}^\lambda d\lambda' \,g(\lambda') 
  \ .
  \label{zeta}
\end{equation}

The level spacing is not studied directly on the $\lambda_i$, because
it depends on the mean level density. To eliminate this dependence,
one ``unfolds'' the spectrum, which means to map the original sequence
$\{\lambda_i\}$ onto a new one $ \{\zeta_i(\lambda_i) \}$ according to
Eq.~(\ref{zeta}) (see e.g. \cite{guhr} for a detailed explanation of
this point). The cumulative spectral function can be expressed in
terms of these new variables:
\begin{equation}
  \hat{\eta}(\zeta) 
  \equiv
  \eta\big(\lambda(\zeta)\big)
  = 
  \zeta + \hat\eta_\mathrm{fluct}(\zeta)
   .
\end{equation}
The distribution of the variable $\zeta$ is uniform in the interval
$[0,1]$ regardless the $g(\lambda)$. 

The nearest-neighbor spacing distribution $P(s)$ gives the probability
that two \emph{neighboring} unfolded eigenvalues $\zeta_i$ and
$\zeta_{i+1}$ are separated by $s$. It is one of the most commonly
used observables in random matrix theory. It is different from the two
level correlation function and it involves all the $k$-level
correlation functions with $k\ge 2$ \cite{mehta}. It displays a high
degree of universality, exhibiting common properties in systems with
very different spectra. Although no general proof has been given, its
shape is thought to depend only on the localization properties of the
states~\cite{guhr}. In the case of the Gaussian Orthogonal Ensemble
(GOE), where all states are extended in the thermodynamic limit, it is
known~\cite{mehta} that $P(s)$, normalized such that $\langle s\rangle
= 1$, follows the so-called \emph{Wigner surmise} (also known as
Wigner-Dyson statistics), namely
\begin{equation}
  P_\mathrm{WD}(s)  = 
  \frac {\pi s}{2} \exp{\left(-\pi s^2/4\right)}  .
  \label{wigner}
\end{equation}
The linear behavior for small $s$ is an expression of the level
repulsion.  This form actually characterizes many different systems
with extended eigenstates (see \textsl{e.g.} ref.~\onlinecite{bohigas}
and references therein). In the case of INM, it has been shown
\cite{deo} that it describes the level spacing better and better as
the fraction of localized states decreases.

On the other hand, a system whose states are all localized will have
completely uncorrelated eigenvalues. This corresponds to a Poisson
process, and the statistics of two adjacent levels is given by
\begin{equation}
  P_\mathrm{P}(s)   =    \exp{(-s)}  .
  \label{poisson}
\end{equation}
If one deals with a set of levels which includes both localized and
extended states, one expects some distribution interpolating between
those two. A natural ansatz is the simple linear combination
\begin{equation}
  P_\mathrm{LC}(s;\pi)   = 
  (1-\pi) P_\mathrm{P}(s) + \pi P_\mathrm{WD}(s) 
  ,
  \label{linear}
\end{equation}
which holds under the hypothesis that contributions coming from
localized and extended modes simply add linearly. Another possibility
comes from a statistical argument due to Wigner (see for
example~\cite{mehta}), that leads to the heuristic function
\begin{equation}
  P(s)   = 
  \mu(s) \exp \left\{ -\int_{0}^{s} ds'\mu(s') \right\} ,
  \label{W}
\end{equation}
where $\mu(s)$ is called \textsl{level repulsion function}. 
Taking $\mu(s)=c_qs^q$, with $q\in[0,1]$, one obtains the Brody
distribution~\cite{brody}
\begin{equation}
  P_\mathrm{B}(s;q)  = 
  c_qs^q\exp{\left( -\frac {c_q s^{q+1}}{q+1}\right)} , 
  \quad
  c_q   = 
  \frac {\Gamma^{q+1}[1/(q+1)]}{q+1} ,
  \label{brody}
\end{equation}
which interpolates between the Poisson ($q=0$) and Wigner-Dyson
($q=1$) distributions. However, this is just another phenomenological
interpolation scheme, since there is no theoretical argument
supporting a level repulsion function increasing as a power law with
an exponent smaller than one.


\section{Method}

\begin{figure}
  \psfrag{r}[][][1.]{$g(\lambda)$}
  \psfrag{l}[][][1.]{$\lambda$}
  \psfrag{z}[][][1.]{$\zeta$}
  \psfrag{l2}[][][.9]{$\lambda$}
  \psfrag{z2}[][][.9]{$\zeta$}
  \psfrag{s/}[][][1.]{$s/\langle s \rangle$}
  \psfrag{P(s)}[][][1.]{$P(s/\langle s \rangle)$}
  \psfrag{D}[][][1.]{$$}
  \psfrag{cumulative}[][][1.2]{}
  \psfrag{zzz}[][][1.]{}
  \psfrag{INM}[][][1.2]{} 
  \psfrag{G11}[][][1.2]{}
  \psfrag{Level}[][][1.2]{}
  \psfrag{s=int}[][][1.]{}
  \centerline{
    \includegraphics[width=.8\columnwidth]{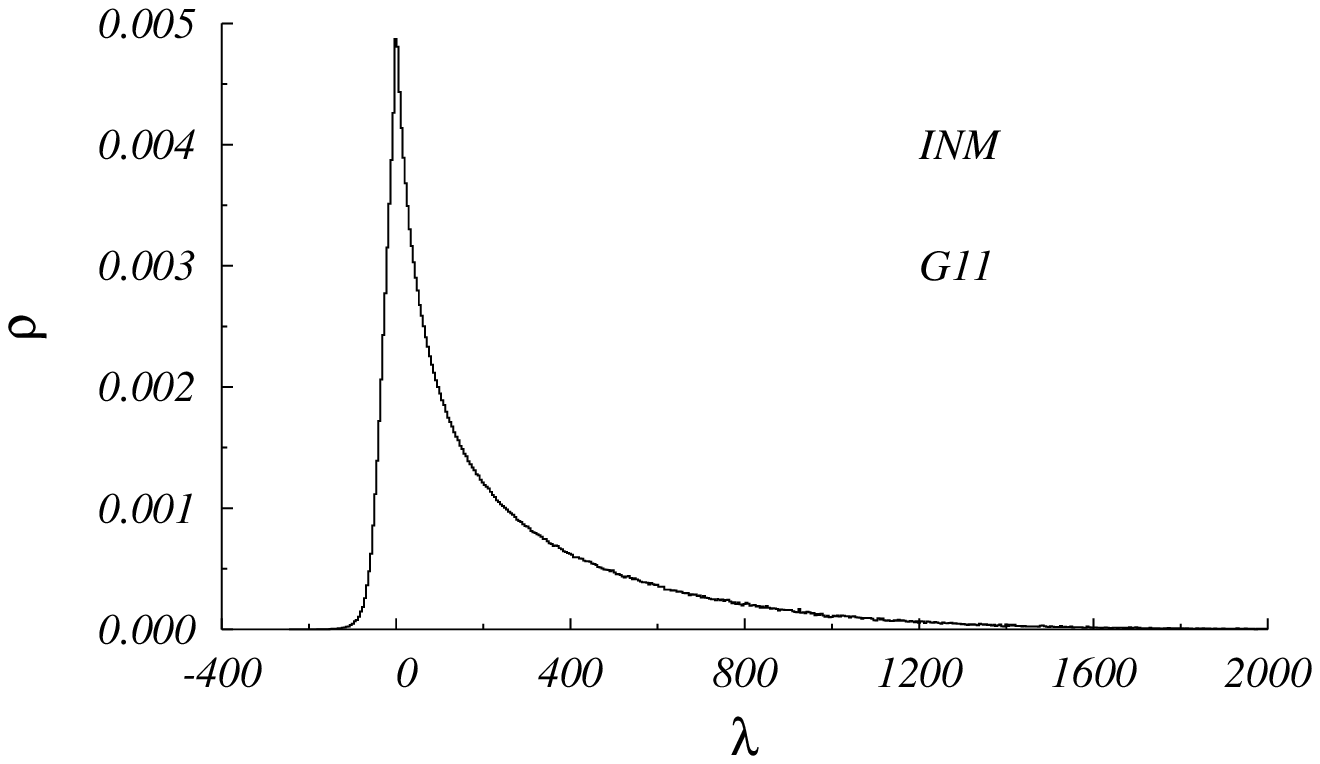}
  }
  \vspace{-.3cm}
  \centerline{
    \includegraphics[width=.8\columnwidth]{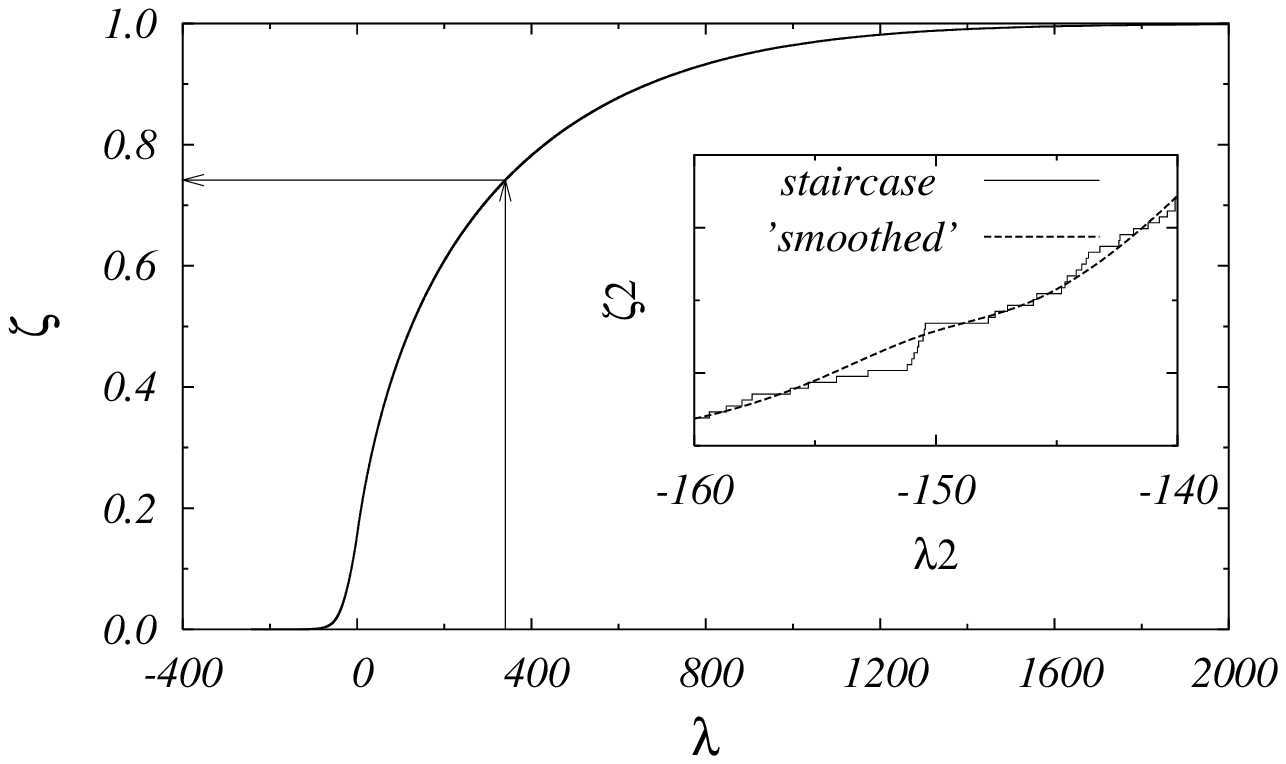}
  }
  \vspace{-.3cm}
  \centerline{
    \includegraphics[width=.8\columnwidth]{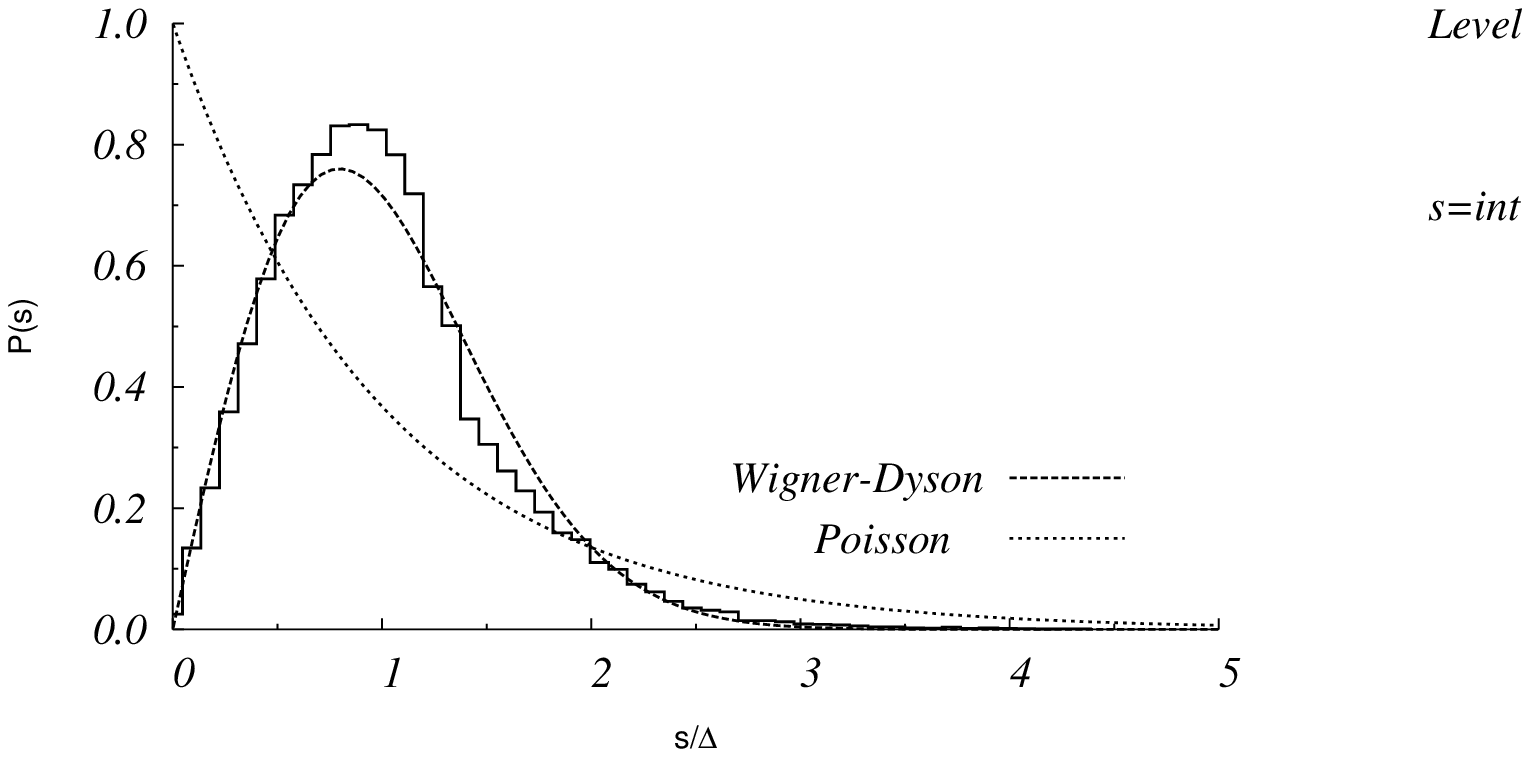}
  }
  \caption{Evaluation of the level-spacing statistics. Top: INM
    spectrum of unit density soft-sphere (pair potential $1/r^12$)
    system at $T=0.68$, as obtained from the numerical diagonalization
    of 100 thermalized configurations. Middle: The cumulative function
    of the same system and decomposition in \textsl{smooth} and
    \textsl{fluctuating} parts (inset). Bottom: The level-spacing
    distribution of this system, normalized to have $\langle s \rangle
    = 1$. Poisson and Wigner-Dyson distributions are also shown for
    comparison.}
  \label{F-unfolding}
\end{figure}

The practical difficulty in performing the unfolding lies in finding a
good approximation to the smooth (averaged) part of the cumulative
spectral function, $\zeta(\lambda)$. We have first obtained a
cumulative function averaged over many samples of the Hessian
(computed from a corresponding number of equilibrium configurations)
and then taken $\zeta(\lambda)$ as the function defined by a cubic
spline interpolation of the resulting staircase. Once this function is
defined, the spacings of each sample can be evaluated by extracting
the $\lambda$ values according to the $g(\lambda)$ and then computing
$s=\zeta(\lambda_{i+1}) -\zeta(\lambda_i)$; the histogram of these
values is an estimate of the $P(s)$. We have also tried digital
filtering (Savistky-Golay \cite{nr}) of the staircase, but the results
were not satisfactory.  The procedure is illustrated in
Fig.~\ref{F-unfolding}.

\begin{figure}
  \psfrag{P1(s)}[][][1]{$P_1(s/\langle s \rangle)$}
  \psfrag{P2(s)}[][][1]{$P_2(s/\langle s \rangle)$}
  \psfrag{s/}[][][1]{$s/\langle s \rangle$}
  \psfrag{D}[][][1]{$$}
  \centerline{
    \includegraphics[width=1.\columnwidth]{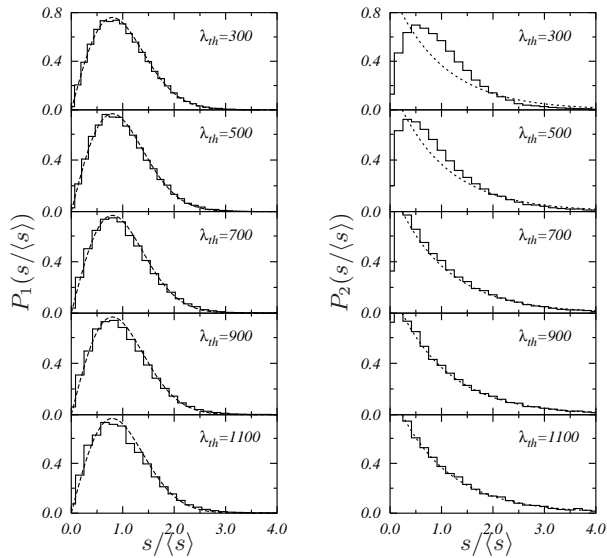}
  }
  \caption{Level spacing distributions $P_1(s/\langle s \rangle)$ and
    $P_2(s/\langle s \rangle)$ obtained from the positive part of the
    spectrum of Fig.~\ref{F-unfolding} for several values of
    $\lambda_\mathrm{th}$. Wigner-Dyson (left) and Poisson (right)
    distributions are also plotted.}
  \label{F-evol_g11}
\end{figure}

To estimate the localization threshold $\lambda_\mathrm{L}$, we
proceed as follows. We divide the full spectrum into two parts at an
arbitrary threshold $\lambda_{th}$ and study (after proper unfolding)
the restricted level-spacing distributions $P_1(s) \equiv P(s |
\lambda<\lambda_{th})$ and $P_2(s) \equiv P(s |
\lambda>\lambda_{th})$.  The localization threshold (where it exists)
should correspond the value of $\lambda_{th}$ that leads to $P_1(s) =
P_\mathrm{WD}(s)$ (extended eigenstates) and $P_2(s) =
P_\mathrm{P}(s)$ (localized eigenstates).  On the other hand if
$\lambda_{th} \neq \lambda_\mathrm{L}$, $P_1(s)$ and $P_2(s)$ will
bear more similarity to each other, since one of them will include
spacings from both localized and extended levels.  A qualitative
feeling of what happens as $\lambda_{th}$ moves through the spectrum
can be gathered from Fig.~\ref{F-evol_g11}, where it can be clearly
seen how $P_2(s)$ evolves from a nearly Wigner to a Poisson
distribution. We remark that these probabilities distributions are
universal since no fitting parameters are required once the plot is
versus $s/\langle s \rangle$, where $\langle s \rangle = \int \!\!
P(s) s \, ds$.


Since one cannot say, based on a finite sample, when one of the
distributions becomes ``exactly'' Poisson or Wigner, one looks for the
value of $\lambda_{th}$ that makes both distributions ``as different
as possible from each other.'' To do this we use (following
ref.~\onlinecite{carpena}) the Jensen-Shannon (JS) divergence as a
measure of the distance between two distributions. It is defined as
\begin{equation}
  D_\mathrm{JS}[P_1,P_2] 
   = 
  H[a_1 P_1+a_2 P_2] -a_1 H[P_1] -a_2 H[P_2]
   .
  \label{djs}
\end{equation}
$H[P]= -\sum_i P(s_i) \log P(s_i)$ is the Shannon entropy of the
distribution $P$, and $a_1,a_2=1-a_1$ are positive weights of
each distribution.  In what follows, we shall choose the weights as
proportional to the support of the section of the (unfolded) spectrum
considered to evaluate the level-spacing. This ensures that the JS
divergence is not affected by differences in sizes~\cite{JS}. The
problem of finding the threshold is then reduced to finding the
maximum of $D_\mathrm{JS}[P_1,P_2]$ as function of $\lambda_{th}$. We
stress that the ideas behind the method are justified only in the
large $N$ limit, and a study of finite-size effects is thus crucial in
this context.


\section{Results}

\subsection{A case study: the GOE}

\begin{figure}
  \psfrag{l}[][][1.]{$\lambda$}
  \psfrag{gl}[][][1]{$g(\lambda)$}
  \centerline{
    \includegraphics[width=.8\columnwidth]{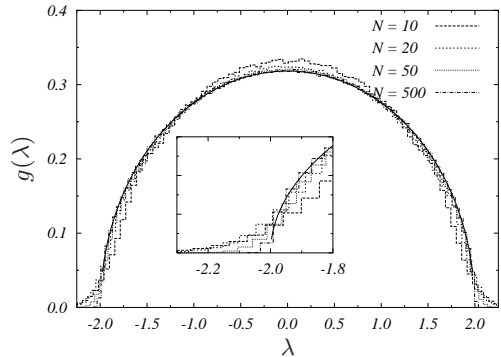}
  }
  \caption{DOS of GOE matrices at several $N$. The solid line
  is the semicircular law predicted at $N\to\infty$. Inset:
  zoom on the left tail.}
  \label{F-dosgoe}
\end{figure}

\begin{figure}
  \psfrag{lth}[][][1]{$\lambda_{th}$}
  \psfrag{q}[][][1.]{$q$}  
  \psfrag{JS}[][][1.]{$D_\text{\tiny JS}\big[P_1,P_2\big]$}
  \psfrag{p}[][][1.]{$\pi$}
  \centerline{
    \includegraphics[width=.8\columnwidth]{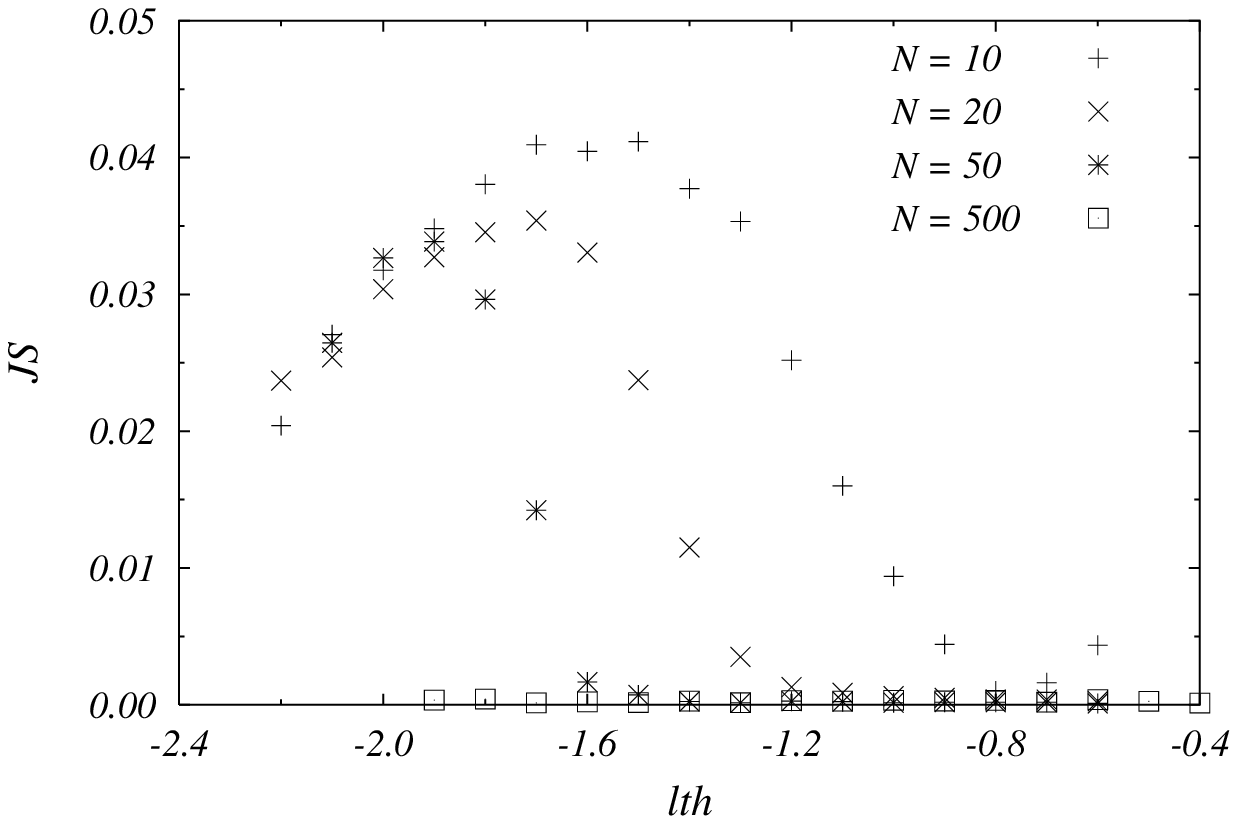}
  }
  \vspace{-.1cm}
  \centerline{
    \includegraphics[width=.8\columnwidth]{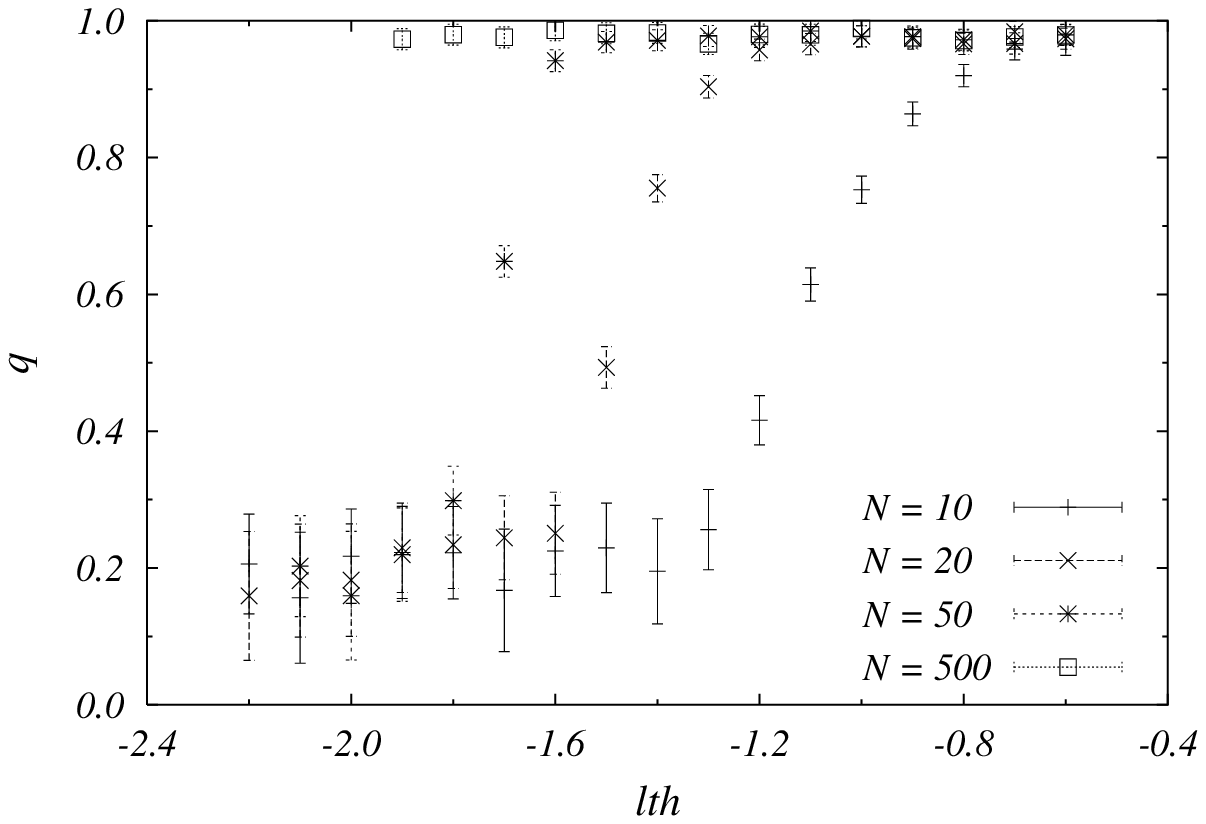}
  }
  \vspace{-.1cm}
  \centerline{
    \includegraphics[width=.8\columnwidth]{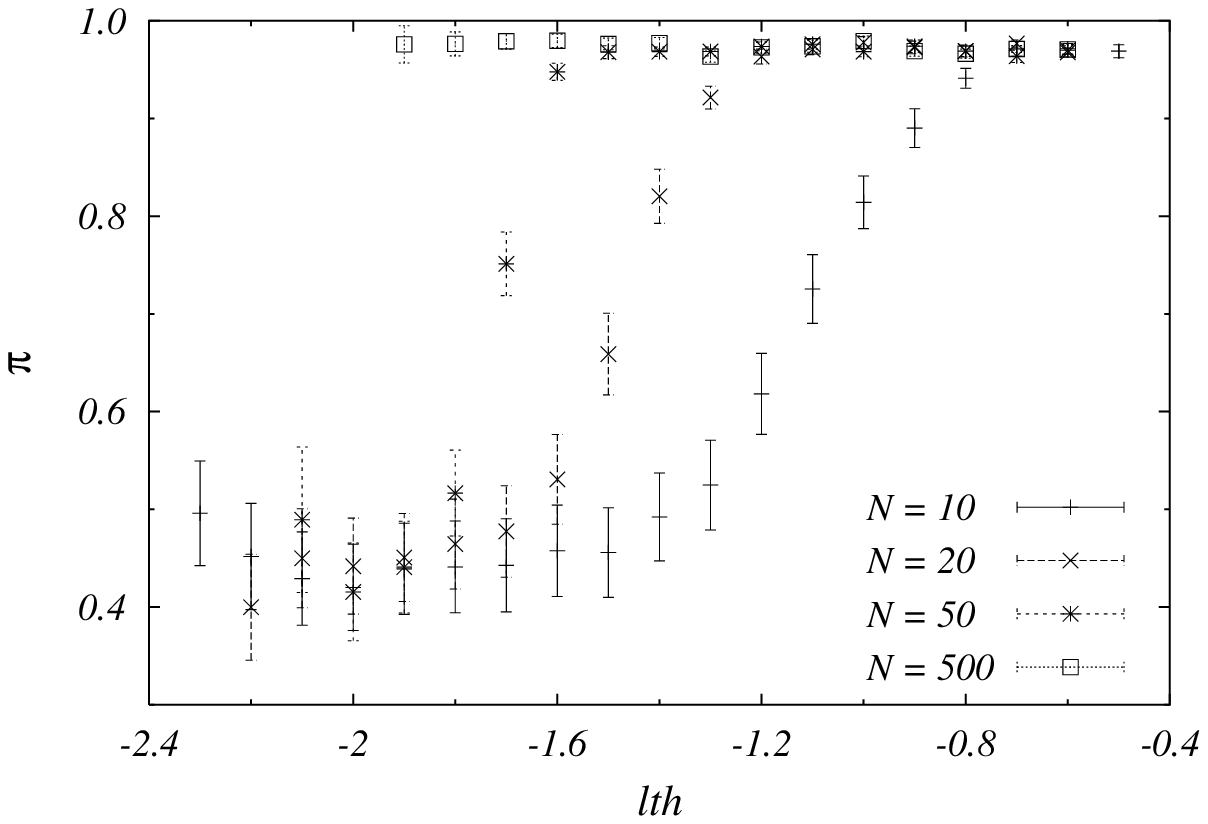}
  }
  \caption{Top: JS divergence for GOE matrices at different $N$. Middle:
  The Brody parameter (see text) for the same values of $N$. Bottom: 
  The order parameter $\pi$ from the linear approximation (\ref{linear}).}
  \label{F-goeJS}
\end{figure}

We have applied this procedure to the GOE, as a test and illustration
of the method. We generated ensembles of $N\times N$ random matrices
for $N=10,20,50,500$ with i.i.d.\ elements (taken from Gaussian
distribution with zero mean and variance $1/\sqrt{N}$) and computed the
DOS by numerical diagonalization (Fig.~\ref{F-dosgoe}). The JS
divergence has a maximum that tends to the band edge as $N$ grows
(Fig.~\ref{F-goeJS}, top), indicating that there is no localization
threshold in this system, as it is known theoretically.

To gain further insight into the workings of the method, we have also
tried fitting the level-spacing distribution restricted to eigenvalues
lower than $\lambda_{th}$ with the functions interpolating between
Poisson and Wigner-Dyson, thus defining a kind of order parameter for
localization ($\pi$ in the case of the linear combination,
Eq.~\ref{linear}, $q$ in the case of the Brody distribution,
Eq.~\ref{brody}). Both $\pi$ and $q$ should be zero if
$\lambda_{th}\leq \lambda_\text{\tiny L}$ and non-zero otherwise. As
Fig.~\ref{F-goeJS} shows, both order parameters start from being
different from their minimum at a value which roughly corresponds to
the maximum of the JS divergence. However, the minimum value is not
zero, most likely due to finite-size problems. Unfortunately, it is
not possible to verify that in this case, because increasing $N$
decreases the fraction of localized states such that their number
remains finite even when $N\to\infty$ \cite{mehta}.

\subsection{The INM spectrum}

We have studied the soft-sphere binary mixture of ref.~\onlinecite{SS}
at unit density and $T=0.2029$ (to be compared with the mode-coupling
critical temperature $T_c\approx 0.2262$). Equilibration of the
supercooled liquid at this temperature has been possible thanks to the
fast Monte Carlo algorithm of ref.~\onlinecite{grpa}. From the
physical point of view, we are interested in studying the nature of
negative modes (which represent about $4.3\%$ of the total modes for
this system). At the temperature considered, the dynamics is highly
arrested, and diffusion events are rare (indeed, at the mean field
level, such as mode-coupling theory, diffusion is completely
suppressed below $T_c$). Accordingly, one expects that all or most of
the negative modes correspond to localized eigenvectors
(\textsl{i.e.\/} local rearrangement of a non extensive number of
particles).  We find that this is not the case.


\begin{figure}
   \psfrag{gl}[][][1.3]{$g(\lambda)$}
   \psfrag{l}[][][1.3]{$\lambda$}
   \centerline{
     \includegraphics[width=\columnwidth]{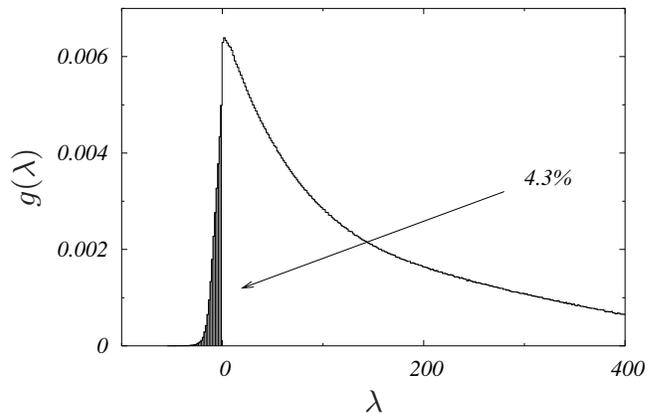}
   }
   \caption{INM spectrum of the binary mixture of soft spheres at
   $\Gamma=1.49$ as obtained from 300 equilibrium configurations
   ($N=2048$).}
   \label{F-dosg149}
\end{figure}


\begin{figure}
   \psfrag{l}[][][1.3]{$\lambda_{th}$}
   \psfrag{gl}[][][1.3]{$g(\lambda)$}
   \psfrag{JS}[][][1.3]{$D_\text{\tiny JS}\big[P_1,P_2\big]$}
  \centerline{
    \includegraphics[width=\columnwidth]{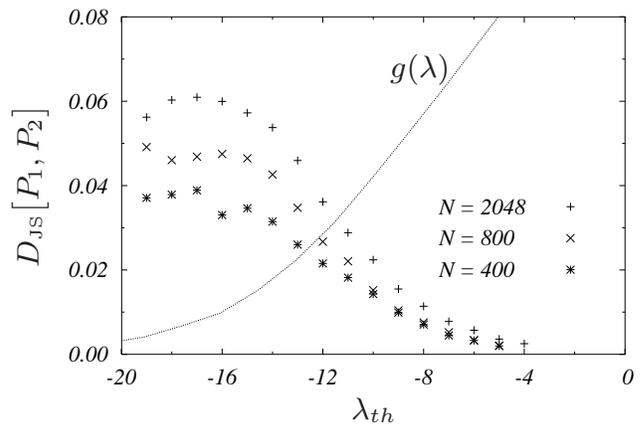}
  }
  \caption{The Jensen-Shannon divergence for the negative tail of the
  spectrum of Fig.~\ref{F-dosg149}. The DOS $g(\lambda)$ is also
  plotted (here it is normalized such that $\int_{-\infty}^0
  g(\lambda) = 1$). }
  \label{F-js149}
\end{figure}


\begin{figure}
  \psfrag{lth}[][][1.3]{$\lambda_{th}$}
  \psfrag{p}[][][1.3]{$\pi$}
  \centerline{
    \includegraphics[width=\columnwidth]{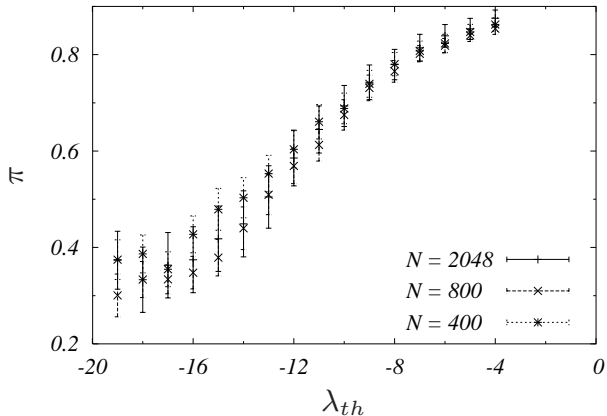}
  }
  \caption{The parameter $\pi$ is found by fitting the level-spacing
  distribution $P(s|\lambda<\lambda_{th})$ to the form in
  (\ref{linear}).}
  \label{F-pi149}
\end{figure}


In Fig.~\ref{F-dosg149} we show the INM spectrum for a system of 2048
soft spheres. The spectrum is expected to have two localization
thresholds (on the positive and negative tails), so to apply the
scheme above we need first to separate the positive and negative
modes. We focus on the negative modes. The JS divergence for the
negative part, evaluated as explained above, is shown in
Fig.~\ref{F-js149} for $N=400,800$ and $2048$ particles. As $N$
increases, the maximum of these curves does not shift as in the GOE
example but it becomes sharper, pointing to a localization
threshold. A quadratic fit of the peak, for the largest size system
leads to $\lambda_\mathrm{L} = -16.8 \pm 1.4$. We also verify that the
two distributions $P_1(s)$ and $P_2(s)$ are indeed Poisson and Wigner
(respectively) for this threshold value.

We next try to fit with the linear interpolation: in
Fig.~\ref{F-pi149} we plot the fitting parameter $\pi$ (cf.\
Eq.~\ref{linear}) for the $P(s|\lambda<\lambda_{th})$. As in the GOE
case, the parameter goes to a non-zero value. At the values of $N$
available to us, there is no clear evidence that larger system sizes
will make $\pi$ go to zero for $\lambda_{th} \le
\lambda_\mathrm{L}$. To check whether this behavior is an indication
of some non-linear effect, we have performed the following
test. Assuming that the threshold is actually at $\lambda_\mathrm{L}$,
for each of the values of $\lambda_{th}$ of Fig.~\ref{F-pi149} we have
generated random spacings distributed with the linear combination of
Poisson and Wigner-Dyson. The weight $\pi$ was taken as proportional
to the number of actual levels between $\lambda_{L}$ and
$\lambda_{th}$, i.e.
\begin{equation*}
  \pi 
  \propto 
  \int_{\lambda_\mathrm{L}}^{\lambda_{th}} g(\lambda) d\lambda 
  \ .
\end{equation*}
We then tried the same fitting procedure we applied to the INM
spectrum, to see whether it would yield the same $\pi$ used. We found
that samples of at least $\approx 90000$ levels were needed in order
to recover the correct weight with the fit (this is more than ten
times greater than the number of levels available from the INM
spectrum of the simulated liquid). Hence we attribute the finite value
of $\pi$ (Fig.~\ref{F-pi149}) to finite-size effects.

\begin{figure}
  \psfrag{l}[][][1.3]{$\lambda_{th}$}
  \psfrag{th}[][][1.3]{$$}
  \psfrag{Brody}[][][1.3]{q}
  \psfrag{JS (bits)}[][][1.3]{$D_\text{\tiny JS}\big[P_1,P_2\big]$}
  \centerline{
    \includegraphics[width=\columnwidth]{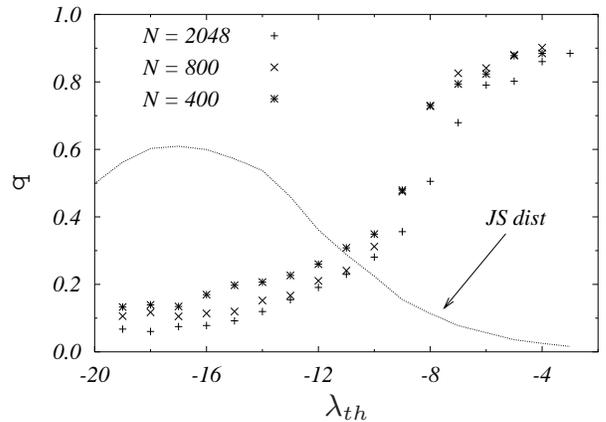}
  }
  \caption{The Brody order parameter for the supercooled liquid at different
  sizes.}
  \label{F-qq149}
\end{figure}

In the fit with the Brody distribution, the finite-size effects seem
to be less pronounced (see Fig.~\ref{F-qq149}). The Brody parameter is
consistent with a localized phase for $\lambda \lesssim -16$: here one
can see that in the large $N$ limit the order parameter goes to zero
as $\lambda \le \lambda_\mathrm{L}$. So the results from the fits
and from the JS divergence are consistent with the existence of a
localization threshold.

Though a more accurate determination of the threshold needs larger
system sizes, this results shows that most (more than 96\%) of the
negative modes in this system are of an extended nature.


\section{Conclusions}

The study of the level-spacing distribution of the INM spectrum of a
glass forming liquid in the supercooled regime we have presented shows
that it is possible to locate the mobility edge for the negative tail
of this spectrum with reasonable precision (other techniques like the
inverse participation ratio are usually less precise). The INM
level-spacing distribution is reasonably described in terms of Wigner
and Poisson distributions and this information can be used to
determine the mobility edge. 

We have applied the technique to the soft-spheres binary liquid below
$T_c$. Our result can be summarized by stating that at this
temperature only $3.4\%$ of the negative modes are localized. This
adds to the evidence (see e.g.\ the critique by Gezelter et
al.~\cite{critique}) that not all extended imaginary modes can be
regarded as leading to free diffusion.  It has been argued \cite{BeLa,
donati00, lanave01, lanave02} that not all extended negative modes
should be considered unstable in the sense of this approach [one
should exclude false saddles (also called shoulder modes) and saddles
that do not connect different minima, an analysis we have not done
here]. Our result implies not only that many negative modes cannot be
regarded as diffusive (not surprising in view of earlier results,
e.g.\ refs.~\onlinecite{donati00,critique}), but also that the vast
majority of these non diffusive negative modes are extended, even below
$T_c$. 

It would be interesting to extend these results to study the
temperature dependence of the localization properties of the INM
across the Mode Coupling temperature. We expect to do this in the near
future.

\section*{Acknowledgments} We thank G.~Biroli, O.~Bohigas, N.~Deo, S.~Franz,  
V.~Mart\'\i{}n-Mayor, G.~Parisi and P.~Verrocchio for useful
discussions and comments. S.C.\ was supported by the ECHP programme
under contract HPRN-CT-2002-00307, {\em DYGLAGEMEM}. T.S.G.\ is a
career scientist of the Consejo Nacional de Investigaciones
Cient\'\i{}ficas y T\'ecnicas (CONICET, Argentina).


\end{document}